\documentclass[%
 aip,
 jmp,%
 amsmath,amssymb,
 reprint,%
]{revtex4-1}

\usepackage{subfigure}
\usepackage{graphicx}% Include figure files
\usepackage{dcolumn}% Align table columns on decimal point
\usepackage{bm}% bold math
\begin{document}

\preprint{AIP/123-QED}

\title{Controlling the efficiency of trapping in treelike fractals}%: A spectrum based approach
% Force line breaks with \\
%\thanks{Footnote to title of article.}

\author{Bin Wu}
\author{Zhongzhi Zhang}
\email{zhangzz@fudan.edu.cn}
\homepage{http://www.researcherid.com/rid/G-5522-2011}

\affiliation {School of Computer Science, Fudan University,
Shanghai 200433, China}

\affiliation {Shanghai Key Lab of Intelligent Information
Processing, Fudan University, Shanghai 200433, China}

\date{\today}% It is always \today, today,

\begin{abstract}
Efficiently controlling the trapping process, especially the trapping efficiency, is central in the study of trap problem in complex systems, since it is a fundamental mechanism for diverse other dynamic processes. Thus, it is of theoretical and practical significance to study the control technique for trapping problem. In this paper, we study the trapping problem in a family of
proposed directed fractals with a deep trap at a central node. The directed fractals are a
generalization of previous undirected fractals by introducing the directed edge weights dominated
by a parameter. We characterize all the eigenvalues and their degeneracies for an associated matrix governing the trapping process. The eigenvalues are provided through an exact recursive relation
deduced from the self-similar structure of the fractals. We also obtain the expressions for the
smallest eigenvalue and the mean first-passage time (MFPT) as a measure of trapping efficiency,
which is the expected time for the walker to first visit the trap. The MFPT is evaluated according to the proved fact that it is approximately equal to reciprocal of the smallest eigenvalue. We show
that the MFPT is controlled by the weight parameter, by modifying which, the MFPT can scale
superlinealy, linearly, or sublinearly with the system size. Thus, this work paves a way to
delicately controlling the trapping process in the fractals.
\end{abstract}

%05.40.Fb Random walks and Levy flights
%61.43.Hv  Fractals; macroscopic aggregates (including diffusion-limited aggregates)
%05.45.Df Fractals
%05.60.Cd Classical transport
%05.40.-a Fluctuation phenomena, random processes, noise, and Brownian motion
%89.75.Hc Networks and genealogical trees
%36.20.¨Cr Macromolecules and polymer molecules

\maketitle

%\tableofcontents

\section{Introduction}

Trapping is a fundamental dynamical process of complex systems, since a large variety of other
dynamical processes occurring in diverse complex systems can be analyzed and understood in terms of
the framework of trapping problem. Examples of these dynamics include light harvesting in antenna
systems~\cite{BaKlKo97,BaKl98,BaKl98JOL,Ag11}, chemical kinetics in molecular
systems~\cite{MoSh58,Ki58,Ko00}, energy or exciton transport in polymer
systems~\cite{BlZu81,SoMaBl97,HeMaKn04,BaKl98JCP}, page search or access in the World Wide
Web~\cite{HwLeKa12,HwLeKa12E}, and so on. All these dynamical processes are closely related to the
trapping process. In view of the direct relevance, it is thus of utmost importance to study
trapping problem in various complex systems.

An interesting quantity related to trapping problem is mean first passage time
(MFPT)~\cite{Re01,NoRi04,CoBeTeVoKl07} defined as the expected time required for a particle to
visit the trap for the first time, which provides a quantitative indicator of trapping
efficiency and gives insight into trapping process. A main theoretical interest in the study of
trapping problem is to understand how the structural properties of the underlying systems affect
the behavior of MFPT. Thus far, MFPT has been intensively studied for trapping in a broad range of
complex networked systems with different structural characteristics~\cite{LiJuZh12}, such as lattices in
different dimensions~\cite{Mo69,GLKo05,GLLiYoEvKo06}, Cayley
trees~\cite{BaKlKo97,BaKl98,BaKl98JOL,BeHoKo03,BeKo06,WuLiZhCh12,LiZh13}, Vicsek
fractals~\cite{WuLiZhCh12,LiZh13} as a model of hyperbranched
polymers~\cite{BlJuKoFe03,BlFeJuKo04,GuBl05,JuvoBe11,FuDoBl13}, treelike
$T-$fractals~\cite{KaRe89,Ag08,HaRo08,LiWuZh10,ZhWuCh11}, Sierpinski
fractals~\cite{KaBa02PRE,KaBa02IJBC,BeTuKo10}, scale-free
fractal~\cite{ZhXiZhGaGu09,ZhXiZhLiGu09,ZhYaGa11,TeBeVo09} or
non-fractal~\cite{KiCaHaAr08,ZhQiZhXiGu09,ZhZhXiChLiGu09,AgBu09,AgBuMa10,MeAgBeVo12,YaZh13}
networks. These previous studies uncovered the discernible influences of different structural
aspects on the trapping efficiency measured by MFPT.

In addition to unveiling the effect of structure on the trapping efficiency, another equally
important target in the study of trapping problem is to control the trapping process, which is
crucial to numerous critical problems. Actually, to drive a networked system towards a desired
function has become an outstanding issue in the area of complex
systems~\cite{LiSlBa11,YaReLaLaLi12,WaNiLaGr12,PoLiSlBa13}. In the context of trapping in complex
systems, it is highly desirable to be able to apply proper control technique to guide the trapping
process with needed trapping efficiency. However, a universal approach for efficiently controlling
trapping process in general complex systems has not been achieved at the present (maybe it does not
exist). Thus, it is of great interest to seek a powerful method steering the trapping process in
specific systems~\cite{BaKlKo97,BaKl98,BaKl98JOL}.

In this paper, we study trapping on a class of treelike regular fractals~\cite{LiWuZh10} with a
deep trap placed at the central node. The fractals being studied include the $T$
fractal~\cite{KaRe86,KaRe89} and the Peano basin fractal~\cite{BaDeVe09} as their two special
cases. We introduce asymmetrical edge weights adjusted by a parameter, which can be used to control
the trapping process through changing the transition probability but without changing the network
structure. Making use of the decimation method~\cite{DoAlBeKa83,CoKa92}, we deduce an exact
recursive relation for the eigenvalues of a matrix governing the trapping process. We then proceed
to find all the eigenvalues and their degeneracies of the relevant matrix. Finally, we provide a
recursive relation for the smallest eigenvalues at two successive generations of the fractals, on
the basis of which we further obtain an approximate expressions of the final smallest eigenvalue
and the MFPT for the trapping problem. We show that, the MFPT can scale as a superinear, linear, or
sublinear function of the system size, depending on the parameter. This work makes it possible by
introducing a method to control the trapping process on fractals towards an ideal case with needed
trapping efficiency.

%%%%%%%%%%%%%%%%%%%%%%%%%%%%%%%%%%%%%%%%%%%%%%%%%%%%%%%%%
% Figure  1
%%%%%%%%%%%%%%%%%%%%%%%%%%%%%%%%%%%%%%%%%%%%%%%%%%%%%%%%%%
\begin{figure}
\begin{center}
\includegraphics[width=0.80\linewidth]{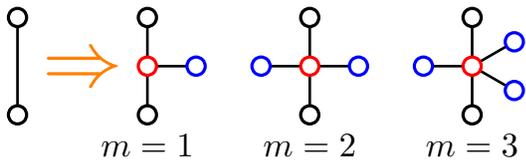}
\caption{ (Color online) Construction of the fractals. The next generation is obtained from current
generation by substituting each edge with the clusters on the right-hand side of the arrow.}
\label{cons}
\end{center}
\end{figure}
%%%%%%%%%%%%%%%%%%%%%%%%%%%%%%%%%%%%%%%%%%%%%%%%%%%%%%%%%%

\section{Constructions and relevant characteristics of the treelike fractals\label{model}}

\begin{figure}
\begin{center}
\includegraphics[width=0.7\linewidth]{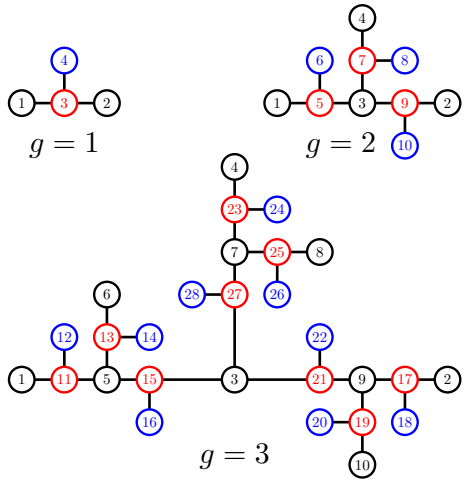}%{G10.eps}
\end{center}
\caption[kurzform]{ (Color online) Iterative growth processes for a special fractal corresponding
to $m=1$.} \label{network}
\end{figure}

The fractal networks under investigation are built in an iterative manner~\cite{LiWuZh10}. Let
$F_{g}$ ($g \ge 0$) represents the fractal graphs after $g$ iterations (generations). For $g=0$,
$F_{0}$ is an edge connecting two nodes. In each following iteration $g \ge 1$, $F_{g}$ is
constructed from $F_{g-1}$ by performing such operations on each existing edge as shown in
Fig.~\ref{cons}: replace the edge by a path two-links long, with both endpoints of the path being
identical to the endpoints of the original edge; then, generate $m$ (a positive integer number) new
nodes and attach each of them to the middle node in the path. Figure~\ref{network} illustrates the
first several iterative processes for a particular fractal corresponding to $m=1$. The fractal
family, parameterized by $m$, subsumes several well-known fractals as its special cases: when $m
=1$, it corresponds to the $T$ fractal~\cite{KaRe86,KaRe89}; when $m = 2$, it is exactly the Peano
basin fractal~\cite{BaDeVe09}.

%%%%%%%%%%%%%%%%%%%%%%%%%%%%%%%%%%%%%%%%%%%%%%%%%%%%%%%%%
% Figure  2
%%%%%%%%%%%%%%%%%%%%%%%%%%%%%%%%%%%%%%%%%%%%%%%%%%%%%%%%%%
%%%%%%%%%%%%%%%%%%%%%%%%%%%%%%%%%%%%%%%%%%%%%%%%%%%%%%%%%%

\begin{figure}
\begin{center}
\includegraphics[width=0.70\linewidth]{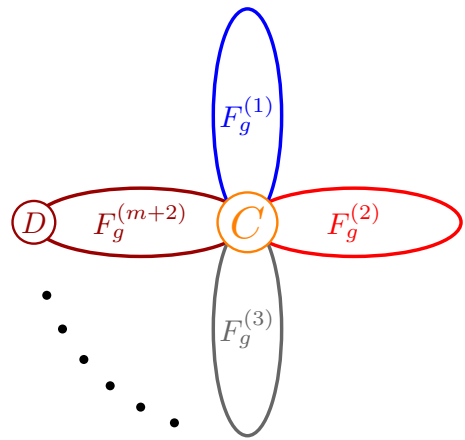}
\end{center}
\caption[kurzform]{(Color online) Another generation approach of the fractals highlighting their
self-similarity. $F_{g+1}$ can be obtained  by joining $m+2$ replicas of $F_{g}$, denoted by
$F_{g}^{(1)}$, $F_{g}^{(2)}$, $\ldots$, $F_{g}^{(m+1)}$, and $F_{g}^{(m+2)}$. $C$ represents the
inmost node, while $D$ denotes an outmost node.}\label{Const2}
\end{figure}

The fractals under consideration are self-similar, which can be seen from the second generation
method. Let us define the central node (e.g., node $3$ in Fig.~\ref{network}) as the inmost node
and those nodes having the largest distance from the central node the outmost nodes, then the
fractals can also be constructed alternatively as follows (see Fig.~\ref{Const2}). Given the
generation $g$, we can obtain $F_{g+1}$ by amalgamating $m+2$ replicas of $F_{g}$ with the $m+2$
outmost nodes in separate duplicates being merged into one single new node, i.e., the inmost node
in $F_{g+1}$.

%%%%%%%%%%%%%%%%%%%%%%%%%%%%%%%%%%%%%%%%%%%%%%%%%%%%%%%%%
% Figure  3
%%%%%%%%%%%%%%%%%%%%%%%%%%%%%%%%%%%%%%%%%%%%%%%%%%%%%%%%%%
%%%%%%%%%%%%%%%%%%%%%%%%%%%%%%%%%%%%%%%%%%%%%%%%%%%%%%%%%%

We can easily derive that the numbers of edges and nodes in $F_{g}$ are $E_{g}=(m+2)^{g}$ and
$N_{g}=E_{g}+1=(m+2)^{g}+1$, respectively. Some relevant features of the fractals can also be
determined~\cite{LiWuZh10,ZhLiZhWuGu09}. Their fractal dimension and random-walk dimension are
separately $d_{\rm f}=\ln (m+2)/\ln2$ and $d_{\rm w}=\ln[2(m+2)]/\ln2$. Therefore, their spectral
dimension is $d_{\rm s}=2d_{\rm f}/d_{\rm w}=2\ln (m+2)/\ln[2(m+2)]$.

\section{Trapping in directed weighted treelike fractals with a single trap at the central node}

The above introduced fractals are a family of important regular fractals, which have received
tremendous attention in the past years~\cite{HaBe87,Fa03,AgViSa09}. A great advantage of regular
fractals is that various problems about them can be treated analytically, deepening the
understanding of geometrical and dynamical behaviors of fractals. In a previous
paper~\cite{LiWuZh10}, we studied the trapping problem in the fractals $F_{g}$ with each edge being
equivalent. In this section, we extend the fractals to directed fractals with special nonnegative
and asymmetrical edge weights, and study trapping problem taking place in them.

\subsection{Relevant definitions for directed weighted treelike fractals}

Let $\vec{F}_{g}$ denote the directed weighted treelike fractals corresponding to $F_{g}$. And let
$W_g$ represent the nonnegative and asymmetrical weight matrix for $\vec{F}_{g}$ such that
$w_{ij}(g)>0$ if and only if there is a directed edge (arc) from node $i$ to node $j$. The weight
of each arc in the directed weighted fractals is defined recursively as follows. When $g=0$,
$\vec{F}_{0}$ has two nodes, denoted by $a$ and $b$, and the weights of arcs $\vec{e}(a,b)$ and
$\vec{e}(b,a)$ are defined to be $W_{ab}(0)=W_{ba}(0)=1$. When $g \geq 1$, by construction, $F_g$
is obtained from $F_{g-1}$ by substituting each undirected edge $e(a,b)$ in $F_{g-1}$ with
two undirected edges $e(a,c)$ and $e(c,b)$, then create $m$ nodes (denoted by $d_1$, $d_2$,
$\cdots$, $d_m$) and attach them to node $c$. The weights of resultant arcs in $\vec{F}_{g}$ are
defined as: $W_{ac}(g)=W_{ab}(g-1)$, $W_{bc}(g)=W_{ba}(g-1)$, $W_{ca}(g)=W_{cb}(g)=1$,
$W_{cd_1}(g)=W_{cd_2}(g)=\cdots=W_{cd_m}(g)=\xi$, and
$W_{d_1c}(g)=W_{d_2c}(g)=\cdots=W_{d_mc}(g)=1$. Here $\xi$ is an arbitrary positive real number,
i.e., $\xi>0$. The parameter $\xi$ is extremely important, since it controls the whole trapping
process.

In undirected weighted networks~\cite{BaBaPaVe04}, node strength is an important quantity
characterizing nodes. Here we extend the definition of strength of a node in undirected weighted
networks to directed weighted networks $\vec{F}_{g}$ by defining the out-strength and in-strength
of node $i$ in $\vec{F}_{g}$ as $s_i^+(g)=\sum_{j=1}^{N_g}W_{ij}(g)$ and
$s_i^-(g)=\sum_{j=1}^{N_g}W_{ji}(g)$, respectively. Furthermore, we define $S_g$ as the diagonal
out-strength matrix of $\vec{F}_{g}$, with the $i$th diagonal entry of $S_g$ being $s_i^+(g)$.

\subsection{Master equation governing the trapping problem}

We now define the trapping problem in the directed weighted treelike fractals $\vec{F}_{g}$ with a
single trap fixed on the central node. Let $t_{ij}(g)=W_{ij}(g)/s_i^+(g)$ denote the probability
per unit time of the particle jumping from node $i$ to its neighboring node $j$. Note that
$t_{ij}(g)$ constitutes an entry of transition matrix $T_g=(S_g)^{-1}W_g$, which suggests that the
powerful tool of the spectral theory~\cite{Ch97} can be employed. Let $x_j(t)$ denote the
probability for the walker being on node $j$ at time $t$. Then,  $x_j(t)$ is governed by the
following master equation~\cite{MoSh58,Ki58,Ko00}
\begin{equation}\label{ME01}
-\frac{{\rm d} x_j(t)}{{\rm d} t}=
x_j(t)-\sum^{N_g}_{\substack{i = 1\\ i \ne {\rm trap}}}t_{ij}(g) x_i(t)\,.
\end{equation}
In this paper, we focus on the a special initial condition $x_j(t=0)=1/(N_g-1)$ for
$j=1,2,\cdots,N_g$ but $j\ne {\rm trap}$.

Equation~(\ref{ME01}) can be recast in the matrix form as
\begin{equation}\label{ME02}
-\frac{d x(t)}{d t}=P_g^{\top}x(t)\,,
\end{equation}
where $x(t)$ is an $N_g-1$ dimensional vector with component $x_i(t)$ ($i=1,2,\cdots,N_g$ but $i\ne
{\rm trap}$), and $P_g$ is an $(N_g-1)\times (N_g-1)$ matrix that is in fact a submatrix of $I_g-T_g$
($I_g$ is the identity matrix) with its row and column corresponding the trap being removed.
Integrating Eq.~(\ref{ME02}) yields
\begin{equation}\label{ME03}
x(t)=e^{-tP_g^{\top}}x(0)\,.
\end{equation}

Evidently, $\vec{F}_{g}$ is strongly connected. Then, the Markov chain  $P_{g}$ is irreducible.
Namely, there exists a unique vector $\phi=(\phi_1,\phi_2,\cdots,\phi_{N_g})^{\top}$ satisfying
that $\phi^{\top}T_{g}=\phi^{\top}$. Moreover, the reciprocal relation $\phi_i t_{ij}(g)=\phi_j
t_{ji}(g)$ holds. Since $P_{g}$ is asymmetric, we introduce a symmetric matrix $Y_g$ with its
$ij$th element being $y_{ij}(g)=\phi_i^{\frac{1}{2}}p_{ij}(g)\phi_j^{-\frac{1}{2}}$, where $p_{ij}(g)$ is the $ij$th entry of $P_g$. By definition,
\begin{equation}\label{ME04}
Y_{g}=I_g-\Phi^{\frac{1}{2}}S_g^{-1} W_g \Phi^{-\frac{1}{2}}
%\Phi^{-\frac{1}{2}}W_g \Phi^{-\frac{1}{2}}
=\Phi^{\frac{1}{2}}P_g \Phi^{-\frac{1}{2}},
\end{equation}
which is real and similar to $P_g$ and thus has the same set of eigenvalues as $P_g$. Here, $\Phi$
is a diagonal matrix with its $i$th diagonal entry being equal to $\phi_i$.

It is easy to check that $P_{g}$ is definitively positive~\cite{Ki58}, thus all its eigenvalues are
positive. Let $\lambda_1(g)$, $\lambda_2(g)$, $\lambda_3(g)$, $\cdots$, $\lambda_{N_g-1}(g)$ be the
$N_g-1$ eigenvalues of matrix $P_{g}$, rearranged as $0<\lambda_1(g)\leq \lambda_2(g) \leq
\lambda_3(g) \leq \ldots \leq \lambda_{N_g-1}(g)$, and let $\psi_1$, $\psi_2$, $\psi_3$, $\ldots$,
$\psi_{N_g-1}$ denote the corresponding normalized, real-valued and mutually orthogonal
eigenvectors. We introduce a more matrix $\Psi $ with its $i$th column vector being Let $\psi_i$.
Then, $Y_g$ can be decomposed as
\begin{equation}\label{ME05}
Y_g=\Psi \Lambda_g \Psi^{-1}=\Psi \Lambda_g \Psi^{\top}\,,
\end{equation}
where $\Lambda_g$ is a diagonal matrix whose $i$th diagonal entry is $\lambda_i(g)$. Thus,
\begin{equation}
P_g^{\top}=(\Phi^{\frac{1}{2}} \Psi) \Lambda (\Phi^{\frac{1}{2}} \Psi)^{-1}\,,
\end{equation}
and Eq.~(\ref{ME03}) can be rewritten as
\begin{eqnarray}
x(t)=(\Phi^{\frac{1}{2}} \Psi) e^{-\Lambda t} (\Phi^{\frac{1}{2}} \Psi)^{-1} x(0)\,.
\end{eqnarray}

For the convenience of description, let $U=\Phi^{\frac{1}{2}} \Psi$ with its $ij$th entry being
denoted by $u_{ij}$, and let $V=(\Phi^{\frac{1}{2}} \Psi)^{-1} x(0)$ be an $N_g-1$ dimensional
vector, the $j$th entry of which is represented by $v_j$. Then,
\begin{eqnarray}\label{ME06}
x_i(t)=\sum_{j=1}^{N_g-1} \exp[-\lambda_j(g) t] u_{ij} v_{j}\,,
\end{eqnarray}
and the survival probability, $c(t)$,  of the walker at time $t$ is
\begin{eqnarray}\label{ME07}
c(t)=\sum_{i=1}^{N_g-1} x_i(t)=\sum_{i=1}^{N_g-1} \sum_{j=1}^{N_g-1} \exp[-\lambda_j(g) t] u_{ij} v_{j}\,.
\end{eqnarray}
Thus, the MFPT, $\langle T \rangle_g$, for trapping in  $\vec{F}_{g}$ with a trap at the central
node is
\begin{equation}\label{ME08}
\langle T \rangle_g = \int_0^{\infty} c(t)dt=\int_0^{\infty}\sum_{i=1}^{N_g-1} \sum_{j=1}^{N_g-1} \exp[-\lambda_j(g) t]
u_{ij} v_{j}dt\,.
\end{equation}
%where $\lim_{t \to +\infty} c(t) t=0$ is used.

If there exists a small $z$ obeying the following condition
\begin{equation}\label{ME09}
\lambda_1(g)=\lambda_2(g)=\cdots =\lambda_z(g) < \lambda_{z+1}(g) \leq \cdots \leq
\lambda_{N_g-1}(g),
\end{equation}
then for sufficient large $t$, we have
\begin{equation}\label{ME10}
c(t)\simeq \exp[-\lambda_1(g) t] \sum_{i=1}^{N_g-1} \sum_{j=1}^{z} u_{ij}v_j\,
\end{equation}
and
\begin{equation}\label{ME11}
\langle T \rangle_g \sim \frac{1}{\lambda_1(g)}\,.
\end{equation}
Below we will show that Eq.~(\ref{ME09}) holds and that $\langle T \rangle_g$ can be evaluated by
Eq.~(\ref{ME11}).

\subsection{Eigenvalues of the related matrix}

After reducing the problem of finding $\langle T \rangle_g$ to determining the minimal eigenvalue
$\lambda_1(g)$ of matrix $P_g$, the next step is to evaluate $\lambda_1(g)$. In what follows we
will use the decimation method~\cite{DoAlBeKa83,CoKa92} to determine the full eigenvalues of matrix
$P_g$. The decimation method makes it possible to solve the eigenvalue problem of $P_g$ of current
iteration based on a similar one from the previous iteration.

We now consider the eigenvalue problem for matrix $P_{g+1}$. Let $\alpha$ denote the set of nodes
belonging to $F_g$, and $\beta$ the set of nodes created at iteration $g+1$. Assume that
$\lambda_{i}(g+1)$ is an eigenvalue of $P_{g+1}$, and $u=(u_{\alpha},u_{\beta})^\top$ is its
corresponding eigenvector, where $u_{\alpha}$ and $u_{\beta}$ correspond to nodes belonging to
$\alpha$ and $\beta$, respectively. Then, eigenvalue equation for matrix $P_{g+1}$  can be
expressed in the following block form:
\begin{equation}\label{T2}
\left[\begin{array}{cccc}
 P_{\alpha,\alpha} & P_{\alpha, \beta} \\
 P_{\beta,\alpha} & P_{\beta, \beta}
\end{array}
\right] \left[\begin{array}{cccc}
 u_{\alpha} \\
 u_{\beta}
\end{array}
\right]=\lambda_{i}(g+1) \left[\begin{array}{cccc}
 u_{\alpha} \\
 u_{\beta}
\end{array}
\right] ,
\end{equation}
where $P_{\alpha,\alpha}$ is the identity matrix, $P_{\beta, \beta}$ is block diagonal with each
block being the same $(m+1)\times (m+1)$ matrix of the form
\begin{equation}\label{TE01}
B=\left[
\begin{array}{ccccc}
 1 & -\frac{\xi}{m\xi+2} & -\frac{\xi}{m\xi+2} & \cdots & -\frac{\xi}{m\xi+2} \\
 -1 & 1 & 0 & \cdots & 0 \\
 -1 & 0 & 1 & \cdots & 0 \\
 \vdots & \vdots & \vdots & \ddots & \vdots \\
 -1 & 0 & 0 & \cdots & 1
\end{array}
\right]\,.
\end{equation}

Eq.~(\ref{T2}) can be rewritten as two equations:
\begin{equation}\label{T3}
P_{\alpha,\alpha}u_{\alpha}+P_{\alpha, \beta}u_{\beta}=\lambda_{i}(g+1)u_{\alpha}\,,
\end{equation}
\begin{equation}\label{T4}
P_{\beta,\alpha}u_{\alpha}+P_{\beta, \beta}u_{\beta}=\lambda_{i}(g+1)u_{\beta}\,.
\end{equation}
Equation~(\ref{T4}) implies
\begin{equation}\label{T4b}
u_{\beta}=[\lambda_{i}(g+1)-P_{\beta,\beta}]^{-1}P_{\beta,\alpha}u_{\alpha}\,,
\end{equation}
provided that the concerned matrix is invertible. Plugging Eq.~(\ref{T4b}) into Eq.~(\ref{T3})
yields
\begin{equation}\label{T5}
\{P_{\alpha,\alpha}+P_{\alpha, \beta}[\lambda_{i}(g+1)-P_{\beta,
\beta}]^{-1}P_{\beta,\alpha}\}u_{\alpha}=\lambda_{i}(g+1)u_{\alpha},
\end{equation}
In this way, the problem of evaluating the eigenvalue $\lambda_{i}(g+1)$ for matrix $P_{g+1}$ is
reduced to determining the eigenvalue problem of a matrix with a smaller order.

Let $Q_g=P_{\alpha,\alpha}+P_{\alpha, \beta}[\lambda_{i}(g+1)-P_{\beta,
\beta}]^{-1}P_{\beta,\alpha}$. In Appendix~\ref{AppA}, we prove that
\begin{equation}\label{T55}
Q_g=(\theta_1+\theta_2)I_g-\theta_2P_g\,,
\end{equation}
where
\begin{equation}\label{TA2}
\theta_1=1+\frac{\lambda_{i}(g+1)-1}{(m\xi+2)[\lambda_{i}(g+1)-2]\lambda_{i}(g+1)+2}
\end{equation}
and
\begin{equation}\label{TA3}
\theta_2=\theta_1-1=\frac{\lambda_{i}(g+1)-1}{(m\xi+2)[\lambda_{i}(g+1)-2]\lambda_{i}(g+1)+2}\,.
\end{equation}
Equation~(\ref{T55}) relates matrix $Q_g$ to matrix $P_g$, which indicates that the eigenvalues for
matrix $P_{g+1}$ can be expressed in terms of eigenvalues for matrix $P_g$.

We next show how obtain the eigenvalues of $P_{g+1}$ through those of $P_g$. According to
Eqs.~(\ref{T5}) and~(\ref{T55}), we have
\begin{equation}\label{A1}
[(\theta_1+\theta_2)I_g -\theta_2 P_g] u_{\alpha} =\lambda_{i}(g+1) u_{\alpha},
\end{equation}
which implies
\begin{equation}\label{A2}
[\theta_1+\theta_2-\lambda_{i}(g+1)]u_{\alpha}=\theta_2 P_g u_{\alpha}\,,
\end{equation}
that is
\begin{equation}\label{A3}
P_g u_{\alpha}=\frac{\theta_1+\theta_2-\lambda_{i}(g+1)}{\theta_2} u_{\alpha}\,.
\end{equation}
Hence, if $\lambda_{i}(g)$ is the eigenvalues of $P_g$ associated with eigenvector $u_a$,
Eq.~(\ref{A3}) means
\begin{equation}\label{A4}
\lambda_{i}(g)=\frac{\theta_1+\theta_2-\lambda_{i}(g+1)}{\theta_2}\,.
\end{equation}
Substituting Eqs.~(\ref{TA2}) and (\ref{TA3}) into Eq.~(\ref{A4}) leads to
\begin{equation}\label{T9}
\lambda_{i}(g)=-(m \xi + 2) [\lambda_{i}(g+1) -2] \lambda_{i}(g+1) \,,
\end{equation}
which can be rewritten as
\begin{equation}\label{T10}
(m \xi + 2)[\lambda_{i}(g+1)]^2-2(m \xi + 2)\lambda_{i}(g+1)+\lambda_{i}(g)=0\,.
\end{equation}
Solving the quadratic equation in the variable $\lambda_{i}(g+1)$ given by Eq.~(\ref{T10}), one
obtains
\begin{equation}\label{T11}
\lambda_{i,1}(g+1)= 1-\sqrt{1-\frac{\lambda_{i}(g)}{m \xi + 2}}
\end{equation}
and
\begin{equation}\label{T12}
\lambda_{i,2}(g+1)=1+\sqrt{1-\frac{\lambda_{i}(g)}{m \xi + 2}} \,.
\end{equation}
Equations~(\ref{T11}) and~(\ref{T12}) relate $\lambda_{i}(g+1)$ to $\lambda_{i}(g)$, with each
eigenvalue $\lambda_{i}(g)$ of $P_{g}$ producing two eigenvalues of $P_{g+1}$. Actually, all eigenvalues of the $P_{g+1}$ can be obtained by these two recursive relations. In Appendix~\ref{AppB}, we determine the multiplicity of each eigenvalue and show that all the eigenvalues can be found by Eqs.~(\ref{T11}) and~(\ref{T12}).

\subsection{Smallest eigenvalue and mean first-passage time\label{MFPT}}

As mentioned above, the smallest eigenvalue of $P_g$ is very important since it is related to the
trapping efficiency $\langle T \rangle_g$: in large systems, the MFPT $\langle T \rangle_g$ for
trapping in $\vec{F}_g$ is proportional to the inverse of the smallest eigenvalue of $P_g$, denoted
by $\lambda_{\rm min}(g)$, that is, $\langle T \rangle_g \sim 1/\lambda_{\rm min}(g)$. Below we
will evaluate $\lambda_{\rm min}(g)$ and $\langle T \rangle_g$, and show how $\langle T \rangle_g$
scales with the system size $N_g$. Before doing this, we first give some useful properties of
eigenvalues for matrix $P_g$.

Let $\Delta_g$ denote the set of the $N_g-1$ eigenvalues of matrix $P_{g}$, that is,
$\Delta_g=\{\lambda_1(g),\lambda_2(g),\lambda_3(g),\cdots, \lambda_{N_g-1}(g)\}$.  On the basis of
above analysis, $\Delta_g$ can be classified into two subsets $\Delta_g^{(1)}$ and $\Delta_g^{(2)}$
such as $\Delta_g=\Delta_g^{(1)} \cup \Delta_g^{(2)}$. $\Delta_g^{(1)}$ contains all eigenvalues 1,
while $\Delta_g^{(2)}$ includes the remain eigenvalues. Thus,
\begin{equation}\label{MFPT01}
\Delta_g^{(1)}= \underbrace {\{ 1,1,1,\ldots,1,1\} }_{m{{(m + 2)}^{g - 1}}}\,.
\end{equation}
These $m(m+2)^{g-1}$ eigenvalues are labeled by $\lambda_{(m+2)^{g-1}+1}(g)$,
$\lambda_{(m+2)^{g-1}+2}(g)$,$\cdots$, $\lambda_{(m+1)(m+2)^{g-1}}(g)$, since they give a natural
increasing order of the eigenvalues for $P_g$, as we will show.

The remaining $2(m+1)^{g-1}$ eigenvalues belonging to $\Delta_g^{(2)}$ are determined by
Eqs.~(\ref{T11}) and~(\ref{T12}). Let $\lambda_1(g-1)$, $\lambda_2(g-1)$, $\lambda_3(g-1)$,
$\cdots$, $\lambda_{(m+2)^{g-1}}(g-1)$ be the $(m+2)^{g-1}$ eigenvalues of matrix $P_{g-1}$,
arranged in an increasing order $\lambda_1(g-1)\leq \lambda_2(g-1) \leq \lambda_3(g-1) \leq \ldots
\leq \lambda_{(m+2)^{g-1}}(g-1)$. Then, for each eigenvalue $\lambda_i(g-1)$ in $P_{g-1}$,
Eqs.~(\ref{T11}) and~(\ref{T12}) generate the two eigenvalues of $P_{g}$, which are labeled as $\lambda_{i}(g)$ and
$\lambda_{(m+2)^{g}-i+1}(g)$:
\begin{equation}\label{MFPT02}
\lambda_{i}(g)= 1-\sqrt{1-\frac{\lambda_{i}(g-1)}{m \xi + 2}}
\end{equation}
and
\begin{equation}\label{MFPT03}
\lambda_{(m+2)^g-i+1}(g)=1+\sqrt{1-\frac{\lambda_{i}(g-1)}{m \xi + 2}} \,.
\end{equation}
Inserting each eigenvalue of $P_{g - 1}$ into Eqs.~(\ref{T11}) and~(\ref{T12}), we can obtain all
the eigenvalues in $\Delta_g^{(2)}$.

It is evident that $\lambda_{i}(g)$ given by
Eq.~(\ref{MFPT02}) monotonously increases with $\lambda_i(g-1)$ and lies in
interval $(0,1)$, while $\lambda_{(m+2)^{g}-i+1}(g)$ provided by Eq.~(\ref{MFPT03}) monotonously decreases with $\lambda_i(g-1)$ and belongs to interval $(1,2)$. Thus, $\lambda_1(g),\lambda_2(g),\lambda_3(g),\cdots,
\lambda_{(m+2)^{g}}(g)$ give an increasing order of all eigenvalues of $P_{g}$.

\begin{figure}
    \begin{center}
        \begin{tabular}{cc}
            \includegraphics[width=80mm]{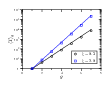} \\
            \includegraphics[width=80mm]{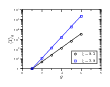} \\
        \end{tabular}
    \caption{Mean first-passage time for the directed weighted fractals with different sizes. The top and below panels correspond to the cases of $m=1$ and $m=2$, respectively. The hollow symbols represents the numerical results, while lines are the predicted results given by Eq.~(\ref{MFPT07}).\label{ATT} }
    \end{center}
\end{figure}

We now begin determine $\lambda_{\rm min}(g)$ and $\langle T \rangle_g$. From above arguments, we
can see that the smallest eigenvalue $\lambda_{\rm min}(g)$ is obtained from $\lambda_{\rm
min}(g-1)$ by using Eq.~(\ref{MFPT02}):
\begin{equation}\label{MFPT04}
\lambda_{\rm min}(g)= 1-\sqrt{1-\frac{\lambda_{\rm min}(g-1)}{m \xi + 2}}
\end{equation}
Using Taylor's formula, we have
\begin{equation}\label{MFPT05}
\lambda_{\rm min}(g) \approx 1-\left[1-\frac{\lambda_{\rm
min}(g-1)}{2(m\xi+2)}\right]=\frac{\lambda_{\rm min}(g-1)}{2(m\xi+2)}\,.
\end{equation}
Considering $\lambda_{\rm min}(1)=1$, Eq.~(\ref{MFPT05}) is solved to yield
\begin{equation}\label{MFPT06}
\lambda_{\rm min}(g) \approx (2m\xi+4)^{1-g}\,.
\end{equation}
Then, the MFPT $\langle T \rangle_g$ for trapping in $\vec{F}_g$ with a trap at the central node is
\begin{equation}\label{MFPT07}
\langle T \rangle_g \sim \frac{1}{\lambda_{\rm min}(g)}=(2m\xi+4)^{g-1}\,.
\end{equation}
In Fig.~\ref{ATT}, we report the numerical and theoretical results of MFPT for the family of directed weighted trees, both of which agree well with each other.

%%%%%%%%%%%%%%%%%%%%%%%%%%%%%%%%%%%%%%%%%%%%%%%%%%%%%%%%%
% Figure  4
%%%%%%%%%%%%%%%%%%%%%%%%%%%%%%%%%%%%%%%%%%%%%%%%%%%%%%%%%%
%%%%%%%%%%%%%%%%%%%%%%%%%%%%%%%%%%%%%%%%%%%%%%%%%%%%%%%%%%

We proceed to express $\langle T \rangle_g$ in terms of the system size $N_{g}$, in order to show
how $\langle T \rangle_g$ behaves with $N_{g}$. From $N_{g}=(m+2)^{g}+1$, we have
$g=\log_{m+2}(N_{g}-1)$. Hence, for very large systems, $\langle T \rangle_g$ can be represented as
a function of $N_g$:
\begin{eqnarray}\label{MFPT08}
\langle T \rangle_{g}\sim (N_{g})^{\ln (2m\xi+4)/ \ln (m+2)}\,.
\end{eqnarray}
When $\xi=1$, Eq.~(\ref{MFPT08}) is consistent with the previously obtained
result~\cite{Ag08,HaRo08,LiWuZh10,ZhWuCh11}.

Equation~(\ref{MFPT08}) shows that for the family of directed weighted fractals $\vec{F}_g$, the
MFPT $\langle T \rangle_g$ scales as a power-law function of the network size $N_{g}$, with the
exponent $\eta(m,\xi)=\ln (2m\xi+4)/ \ln (m+2)$ depending on $m$ and $\xi$. When $\xi>(m-2)/(2m)$,
$\langle T \rangle_g$ varies superlinearly with $N_{g}$; when $\xi=(m-2)/(2m)$, $\langle T
\rangle_g$ scales linearly with $N_{g}$; and when $\xi<(m-2)/(2m)$, $\langle T \rangle_g$ behaves
sublinearly with $N_{g}$. Thus, the MFPT $\langle T \rangle_g$ displays rich behavior by changing
$\xi$ to control (adjust) transition probability: when $\xi$ decreases from infinite to zero, the
trapping efficiency $\langle T \rangle_g$ covers a range from superlinear dependence (less
efficient trapping) to sublinear dependence (highly efficient trapping) on the network size $N_{g}$.

\section{Conclusions}

By introducing asymmetrical positive edge weights  controlled by a parameter, we have presented a class of
directed weighted treelike fractals. We have studied the trapping problem on the directed fractals
with a perfect trap positioned at the central node. According to the self-similar structure of the
fractals, we have characterized all the eigenvalues and their multiplicities of a relevant matrix
representing the random walk rate equation. The eigenvalues were deduced from a recursive relation
governing eigenvalues at two successive iterations of the directed fractals. We have
also studied the properties of the eigenvalues, based on which we have derived a recursion
expression between the smallest eigenvalues of the fractals at two consecutive generations and
obtained an approximate analytical result for the smallest eigenvalue.

Furthermore, according to the obtained fact that the MFPT to the trap approximately equals the
inverse of the smallest eigenvalue, we have computed the MFPT. The result shows that by tuning the
weight parameter, the MFPT exhibits rich behavior, which can scales superlinearly, linearly, or
sublinealy with the system size, depending on the weight parameter. Thus, by changing the weight
parameter, we can control the trapping process in the fractal systems in order to have needed trapping
efficiency. We expect that the introduced weights can also be used to tailor the systems to carry
out other desirable functions as wanted.

\begin{acknowledgments}
This work was supported by the National Natural Science Foundation
of China under Grant Nos. 61074119 and 11275049.
\end{acknowledgments}

\appendix

\section{Proof of equation~(\ref{T55}) \label{AppA}}

Since $Q_g=P_{\alpha,\alpha}+P_{\alpha, \beta}[\lambda_{i}(g+1)-P_{\beta,
\beta}]^{-1}P_{\beta,\alpha}$ is relevant to the inverse of matrix $\lambda_{i}(g+1)-P_{\beta,
\beta}$ that is block diagonal, then $[\lambda_{i}(g+1)-P_{\beta, \beta}]^{-1}$ is also block
diagonal with each block being $[\lambda_{i}(g+1)-B]^{-1}$. In order to prove Eq.~(\ref{T55}), we
use $H=(h_{ij})_{(m+1)\times (m+1)}$ to denote $[\lambda_{i}(g+1)-B]^{-1}$, and rewrite
$P_{\alpha,\beta}$ and $P_{\beta,\alpha}$ as
\begin{equation}\label{App01}
P_{\alpha,\beta}=(U_1,U_2,\cdots,U_{E_g})\,
\end{equation}
and
\begin{equation}\label{App02}
P_{\beta,\alpha}=\left(
\begin{array}{c}
V_1 \\
V_2 \\
\vdots \\
V_{E_g} \\
\end{array}\right)\,,
\end{equation}
respectively. In Eqs.~(\ref{App01}) and~(\ref{App02}), $E_g$ denotes the number of edges in $F_g$;
$U_i$ ($1 \leq i \leq E_g$) is a matrix of order $(m+2)^g \times (m+1)$ describing the transition
rate from nodes of $F_g$ to $m+1$ nodes newly generated by the $i$th edge of $F_g$; similarly,
$V_i$ ($1 \leq i \leq E_g$) is an $(m+1)\times (m+2)^g$ matrix indicating the transition
probability from the $m+1$ new nodes created by the $i$th edge to those old nodes belonging to
$F_g$. Thus $Q_g$ can be expressed as
\begin{eqnarray}\label{HA1}
Q_g&=&P_{\alpha,\alpha}+P_{\alpha, \beta}[\lambda_{i}(g+1)-P_{\beta,
\beta}]^{-1}P_{\beta,\alpha}\nonumber\\
&=&I_g+\sum_{i=1}^{E_g} U_i\, H \,V_i \nonumber\\
&=&I_g+\sum_{i=1}^{E_g} \left(a_i \varepsilon_{l_i}+b_i \varepsilon_{r_i}, 0, \cdots, 0\right) H
\left(\begin{array}{c}
-\frac{\varepsilon_{l_i}^{\top} + \varepsilon_{r_i}^{\top}}{m\xi+2} \\
0 \\
\vdots \\
0 \\
\end{array}\right)\nonumber\\
&=&I_g+\sum_{i=1}^{E_g} h_{11} (a_i \varepsilon_{l_i}+b_i \varepsilon_{r_i})\left(-\frac{\varepsilon_{l_i}^{\top} + \varepsilon_{r_i}^{\top}}{m\xi+2}\right)\nonumber\\
&=&I_g-\frac{h_{11}}{m\xi+2}\nonumber\\
&&\times \sum_{i=1}^{E_g} \left(a_i \varepsilon_{l_i} \varepsilon_{l_i}^{\top}+ a_i \varepsilon_{l_i} \varepsilon_{r_i}^{\top} + b_i \varepsilon_{r_i} \varepsilon_{l_i}^{\top} + b_i \varepsilon_{r_i} \varepsilon_{r_i}^{\top}\right)\nonumber\\
&=&I_g-\frac{h_{11}}{m\xi+2} (P_g-2I_g)\,,
\end{eqnarray}
where $l_i$ and $r_i$ are the two endpoints of the $i$th edge of $F_g$; $\varepsilon_i$ is
a vector having only one nonzero element $1$ at $i$th entry with other entries being zeros; $a_i$ and $b_i$ are two entries of
$P_g$ corresponding to directed edges $(l_i,r_i)$ and $(r_i,l_i)$, respectively.

In order to prove Eq.~(\ref{T55}), the only thing left is to determine $h_{11}$, which is the entry
at the first row and first column of matrix $[\lambda_{i}(g+1)-B]^{-1}$. By definition,
\begin{equation}\label{H1}
[\lambda_{i}(g+1)-B]^{-1}=\frac{[\lambda_{i}(g+1)-B]^{*}}{{\rm det}[\lambda_{i}(g+1)-B]}\,,
\end{equation}
where $[\lambda_{i}(g+1)-B]^{*}$ is the complex adjugate matrix of $\lambda_{i}(g+1)-B$. It is easy
to verify that
\begin{small}
\begin{equation}\label{H2}
{\rm
det}[\lambda_{i}(g+1)-B]=[\lambda_{i}(g+1)-1]^{m+1}-\frac{m\xi}{m\xi+2}[\lambda_{i}(g+1)-1]^{m-1}\,.
\end{equation}
\end{small}
Then, we have
\begin{eqnarray}\label{H3}
h_{11}&=&\frac{[\lambda_{i}(g+1)-1]^m}{{\rm
det}[\lambda_{i}(g+1)-B]}\nonumber\\
&=&\frac{(m\xi+2)[\lambda_{i}(g+1)-1]}{(m\xi+2)[\lambda_{i}(g+1)-2]\lambda_{i}(g+1)+2}\nonumber\\
&=&(m\xi+2)\theta_2\,,
\end{eqnarray}
Inserting Eq.~(\ref{H3}) into (\ref{HA1}) and considering $\theta_2=\theta_1-1$ yields
\begin{equation}
Q_g=I_g-\theta_2(P_g-2I_g)=(\theta_1+\theta_2)I_g-\theta_2 P_g\,.
\end{equation}
This completes the proof of Eq.~(\ref{T55}).

\section{Multiplicities of eigenvalues \label{AppB}}

By computing the eigenvalues numerically, one can find some important properties about the
eigenvalues. First, all eigenvalues appearing at a given iteration $g_{i}$ continue to appear at
all subsequent generations greater than $g_{i}$. Second, all new eigenvalues appearing at iteration
$g_{i}+1$ are just those generated via Eqs.~(\ref{T11}) and~(\ref{T12}) by substituting
$\lambda_{i}(g)$ with $\lambda_{i}(g_{i})$ that are newly created at iteration $g_{i}$. Namely, all
eigenvalues can be obtained by Eqs.~(\ref{T11}) and~(\ref{T12}). Thus, all that is left is to
determine the multiplicities of the eigenvalues, on the basis of the two fundamental natures of the
eigenvalues.

Let $M_g(\lambda)$ denote the multiplicity of eigenvalue $\lambda$ of matrix $P_g$. Since all
eigenvalues are generated from eigenvalue $1$, we first determine the number of eigenvalue $1$ for
$P_g$. To this end, let $r(X)$ denote the rank of matrix $X$. Then
\begin{equation}\label{N0}
M_g(\lambda=1)=(m+2)^g-r(P_g-1\times I_g)\,.
\end{equation}
When $g=1$, it is obvious that $M_1(\lambda=1)=m+2$. When $g=2$, $P_g-I_g$ is block diagonal, with
each of its $m+2$ blocks having the same form as
\begin{equation}\label{N1}
\left[\begin{array}{ccccc}
 0 & -\frac{1}{m\xi+2} & -\frac{\xi}{m\xi+2} & \cdots & -\frac{\xi}{m\xi+2} \\
 -1 & 0 & 0 & \cdots & 0 \\
 -1 & 0 & 0 & \cdots & 0 \\
 \vdots & \vdots & \vdots & \ddots & \vdots \\
 -1 & 0 & 0 & \cdots & 0
\end{array}
\right]_{(m+2)\times(m+2)}\,.
\end{equation}
Since the rank of each block is 2, we have $M_2(\lambda=1)=(m+2)^2-2 (m+2)=m(m+2)$.

We continue to determine $M_g(\lambda=1)$ for $g>2$. For this purpose, we consider another case of
trapping in $\vec{F}_g$ with the trap located at an initial node, i.e., a node belonging to $F_0$.
For this case of trapping, we introduce matrix $B_g$, which is the counterpart  $P_g$ of the case
that the central node is the trap. In addition, we define $A_g = B_g-I_g$. Then, for $g \geqslant
2$, $P_{g+1}-I_{g+1}$ can be expressed in terms of $A_g$:
\begin{equation}\label{N2}
P_{g+1}-I_{g+1}=\left[
\begin{array}{ccccc}
A_{g} & 0 & 0 & \cdots & 0 \\
0 & A_{g} & 0 & \cdots & 0 \\
0 & 0 & A_{g} & \cdots & 0 \\
\vdots & \vdots & \vdots & \ddots & \vdots \\
0 & 0 & 0 & \cdots & A_{g}
\end{array}
\right]\,,
\end{equation}
with $A_g$ ($g \geqslant 2$) obeying
\begin{equation}\label{N3}
A_{g}=\left[
\begin{array}{ccccc}
A_{g-1} & 0 & 0 & \cdots & -\mu_1 \\
0 & A_{g-1} & 0 & \cdots & -\mu_2 \\
0 & 0 & A_{g-1} & \cdots & -\mu_3 \\
\vdots & \vdots & \vdots & \ddots & \vdots \\
-\mu_1^{\top} & -w_2^{\top} & -w_3^{\top} & \cdots & A_{g-1}
\end{array}
\right]\,,
\end{equation}
in which each $\mu_i$ ($1 \leq i \leq m+1$) is an $(m+2)^{g-1}\times (m+2)^{g-1}$ matrix that has a
unique nonzero entry $1/(m\xi+2)$ describing the transition probability from a node in one copy of
$F_g^{(i)}$ to the inmost node being amalgamated to form $F_{g+1}$; each $w_i$ ($2 \leq i \leq
m+1$) is an $(m+2)^{g-1}\times (m+2)^{g-1}$ matrix, which has only one nonzero element
$\xi/(m\xi+2)$ indicating the transition probability from the inmost node to one node in a replica
of $F_g^{(i)}$.

Note that for any neighbor $u$ of the central node in $\vec{F}_g$ ($g \geq 2$), it has a neighbor
$h$ with both the in-degree and the out-degree being $1$. Thus, in matrix $A_g$, there is only one
nonzero entry at row $h$ and column $h$, respectively, that is, $(h,u)$ and $(u,h)$. Hence, by
using some basic operations for matrix, we can eliminate all nonzero elements at the last row and
column of $A_g$. Then, we have $r(A_g)=r(P_g-I_g)$, implying $r(P_{g+1}-I_{g+1})=(m+2)r(P_g-I_g)$.
Considering $r(P_2-I_2)=2(m+2)$, we obtain $r(P_g-I_g)=2(m+2)^{g-1}$. Thus the multiplicity of
eigenvalue $1$ is
\begin{equation}\label{N4}
M_g(\lambda=1)=\begin{cases}
m+2, &g=1, \\
m(m+2)^{g-1}, &g \geqslant 2.
\end{cases}
\end{equation}

We hasten to compute the multiplicities of other eigenvalues. As mentioned above, every other
eigenvalue in $P_g$ is generated from eigenvalue $1$ (i.e., a descendant of eigenvalue 1) and keeps
the multiplicity of its father. Then, the multiplicity of each first-generation descendant of
eigenvalue 1 is $m(m+2)^{g-2}$, the multiplicity of each second-generation descendant of eigenvalue
1 is $m(m+2)^{g-3}$, the multiplicity of each $(g-2)$nd generation descendant of eigenvalue 1 is
$m(m+2)$, and the multiplicity of each $(g-1)$st generation descendant of eigenvalue 1 is $m+2$.
Moreover, it is easy to check that the number of the $i$th ($0\leq i\leq g-1$) generation distinct
descendants of eigenvalue 1 is $2^{i}$, where $0$th generation descendants mean eigenvalues 1
themselves. Thus, the total number of eigenvalues of $P_g$ is found to be
\begin{equation}
\sum_{i=0}^{g-2}m(m+2)^{g-1-i}2^{i}+(m+2)2^{g-1}= (m+2)^g\,,
\end{equation}
implying that all the eigenvalues of $P_g$ are successfully found.

\nocite{*}
%\bibliography{aipsamp}% Produces the bibliography via BibTeX.

\end{document}